\documentclass{elsart3}

\usepackage{graphicx}
\usepackage{amssymb}

\newcommand{\ket}[1]{|{#1}\rangle{}}

\newcommand{\beq}{\begin{equation}}
\newcommand{\eeq}{\end{equation}}

\begin{document}

\begin{frontmatter}



\title{Exotic quantum phases and phase transitions in correlated
  matter\thanksref{label0}}
\thanks[label0]{This paper was written by Fabien Alet and Aleksandra Walczak with equal contribution and in consultation with Matthew Fisher, based on his lectures and transparencies.}


\author[label1,label2]{Fabien Alet},
\author[label3]{Aleksandra M. Walczak} and
\author[label4]{Matthew P. A. Fisher}
\address[label1]{Laboratoire de Physique Th\'eorique, UMR 5152 of CNRS, Universit\'e Paul Sabatier, 31062 Toulouse, France}
\address[label2]{Service de Physique Th\'eorique, URA 2306 of CNRS, CEA Saclay, 91191 Gif sur Yvette, France}
\address[label3]{Department of Physics and Center for Theoretical Biological
  Physics, University of California at San Diego, La Jolla, California
  92093, USA}
\address[label4]{Kavli Institute for Theoretical Physics, University of California, Santa Barbara, California 93106, USA}

\begin{abstract}
We present a pedagogical overview of recent theoretical work on
unconventional quantum phases and quantum phase transitions in condensed matter systems. Strong correlations
between electrons can lead to a breakdown of two traditional paradigms of
solid state physics: Landau's theories of Fermi liquids and phase
transitions. We discuss two resulting ``exotic'' states of matter:
topological and critical spin liquids. These two quantum phases do not
display any long-range order even at zero temperature. In each case, we show how a gauge theory description is
useful to describe the new concepts of topological order, fractionalization
and deconfinement of excitations which can be present in such spin liquids. We make brief connections,
when possible, to experiments in which the corresponding physics can be
probed. Finally, we review recent work on deconfined quantum critical
points. The tone of these lecture notes is expository: focus is on gaining a
physical picture and understanding, with technical details kept to a minimum.
\end{abstract}

\end{frontmatter}

% main text
\section{Introduction}

Traditional condensed matter physics is shaped by two ideas that can be
traced back to Landau: the Fermi liquid theory and the theory of phase transitions
\cite{textbook}. These principles have proven very fruitful and enabled
great progress in describing simple metals, semiconductors and
insulators. However starting from the experimental discovery of the
fractional quantum Hall effect \cite{QHE}, and before that theoretical
predictions in $1d$ quantum spin chains \cite{QSC}, it has become clear that
a large number of phenomena observed in condensed matter materials lie
beyond the description of these two fundamental paradigms. Such effects are
observed in systems with strongly correlated electrons, which include for
instance high $T_c$ superconductors \cite{SUP}, heavy fermion materials \cite{HFM} or
transition metal oxides \cite{TMO}. In these systems the interactions
between the electrons need to be considered explicitly and are primarily
responsible for the observed phases of matter, which cannot be explained in
a Fermi liquid picture. These new quantum
mechanical phases are present at zero temperature, but their effects can
also be felt at higher temperatures. Furthermore the phases of matter which
emerge do not necessarily differ by symmetry, although 
quantum phase transitions between them are sometimes possible: this is not
reconciliable within the Landau theory of phase transitions. Therefore the standard
Landau paradigms do not apply and one must seek new approaches to understand
the nature of these exotic phases.

These notes are based on the lectures entitled ``Exotic quantum phases and
phase transitions in correlated matter'' given by one of us (M.P.A.F) at the FPSPXI summer school
in Leuven. They are meant to be pedagogical and they also reflect the
process of understanding of the scribes (F.A. and A.M.W.). The lecture notes can therefore
be useful as a basic introduction to the topic. We have also tried to keep in the tone of these notes the lively spirit of the lectures. 

The discussion will consider one of the simplest systems in which strongly
correlated electron physics emerges, that of Mott insulators and will focus
in particular on the example of spin liquid phases. Spin liquids are exotic phases that
occur
in quantum spin systems with odd number of spins $S=1/2$ per unit cell when no symmetries
are broken at  zero temperature; as such, they can be
thought of as unusual quantum paramagnets. We start with a discussion in
Sec.~\ref{sec:break} of the situations where the interactions between
electrons matter and how these strong correlations can 
break down the traditional solid state physics paradigms. In
the following two sections, we present two known classes of systems where these
strong interactions stabilize exotic phases: topological (Sec.~\ref{sec:topo}) and
critical or algebraic (Sec.~\ref{sec:crit}) spin liquids. We then present in
Sec.~\ref{sec:dec} recent work on an explicit example of a quantum phase
transition which cannot be explained naturally in terms of a Landau theory,
and we conclude in Sec.~\ref{sec:conc}.

\section{Breakdown of traditional paradigms}
\label{sec:break}
\subsection{Fermi liquid theory and strong correlations}

Fermi liquid theory has been remarkably successful explaining why band
theory works, even in systems with interacting
electrons. Band theory basically assumes free electrons. Assuming a one-to-one correspondence between the states of the free
system and those of the interacting one, the Fermi liquid theory shows that
interactions essentially dress up the electrons on the Fermi surface into a
``quasi-particle'' with admittedly renormalized characteristics (effective
mass etc.), but with retained independent properties - justifying band
theory. 

According to the resulting band theory, 
electrons have allowed energy levels (bands) separated by forbidden energy regions
(gaps) in the presence of the ionic periodic potential. For temperatures smaller than needed
to thermally excite an electron over the gap, the electrons cannot move around if the highest energy band is
filled and consequently the material is an insulator,
whereas if the highest band is partially filled the material is a
metal. Effectively, band theory predicts that systems with an even
number of electrons per unit cell are usually insulators and those with an odd
number are always metals. 

These predictions are correct if the bands are wide enough, as
compared to the characteristic Coulomb interaction between electrons,
such that the Fermi liquid assumption holds. This is usually the case for Fermi
surfaces formed by electrons with an atomic $s$ or $p$ character.
For materials with electrons at the Fermi energy from partially filled atomic
$d$ or $f$ orbitals, the overlap of these orbitals on neighbouring atoms is usually much
smaller, since the interatomic spacing is still determined by the $s$ and
$p$ orbitals. This results in a smaller kinetic energy of hopping and
therefore narrower bands. In such cases the electrons are said to be {\it strongly correlated}. 
It is worth stressing that the Coulomb energy has the same magnitude in either case - it
is the kinetic energy of the electrons which differs between systems of
strongly correlated electrons and those correctly described by Fermi liquid
theory. The validity of Fermi liquid theory is determined by the ratio of Coulomb and kinetic energies.

Strong interactions lead to a correlated motion of electrons, and
understanding its precise nature is the challenge faced by theories of strongly correlated electron systems.
For larger magnitudes of Coulomb repulsion, the electrons are less likely to hop to
a neighbouring site and they effectively self-localize on a given lattice site. To
be more explicit, the electrons can virtually hop to another site and back,
however the materials are insulators and will not conduct. Such materials, 
with an odd number of electrons per unit cell, that do no conduct because
of Coulomb repulsion are named {\it Mott
insulators}, as opposed to band insulators which result from filled energy
bands. In fact band theory, which ignores the correlation effects
that determine the nature of these materials, predicts that they are
metals. Examples of such materials are transition metal oxides
(e.g. cuprates, manganites, chlorides) with partially filled $3d$ and $4d$
bands and rare earth and heavy fermion materials with partially filled $4f$
and $5f$ bands.   

The simplest model which takes into account the correlation effects is the
Hubbard model
\beq
H=-t \sum_{\left< ij\right> ,\alpha}[ c^{\dagger}_{i,\alpha}c_{j,\alpha} + h.c.] + U
\sum_i n_{i,\uparrow}n_{i,\downarrow},
\label{HubHam}
\eeq

in which the inter-site hopping of the electrons $t$ competes with on-site 
repulsion $U$. Here $c^{\dagger}_i, c_i, n_i$ are respectively
creation, annihilation and density operators of an electron at lattice site
$i$ with spin index $\alpha$ which can take on up or down values
($\alpha=\uparrow ,\downarrow$). The model considers one electron per
cell on a lattice and $\left< ij\right>$ denotes pairs of neighbouring sites $i$ and
$j$. If $t/U$ is large, electrons are essentially free. In the other limit, $U/t\gg 1$, the electrons are site localized and the
nature of the system is shaped by the residual spin physics. 

Kinetic energy favours two neighboring site-localized electrons to have
antialigned spins, as a hop to a neighbouring site
with the same spin orientation is not allowed due to the Pauli exclusion principle.
Therefore at half-filling and in the large repulsion limit, the most elementary model that captures
the dominant interactions in Mott insulators is the $S=1/2$ Heisenberg
Hamiltonian with antiferromagnetic exchange $J$
\begin{equation}
H=J \sum_{\left< ij\right>}{\bf S_i}\cdot {\bf S_j}.
\label{HeisHam0}
\end{equation}
Here ${\bf S}=(S^x,S^y,S^z)$ are $S=1/2$ operators, which (up to a
constant) are equal to the Pauli matrices $(\sigma_x,\sigma_y,\sigma_z)$. The strength of the
exchange constant is determined in terms of the original energy scales by a second order virtual hopping process: $J\sim t^2 / U$.

Although these models are quite simple to write down, they prove very complicated to solve. The
Hubbard model, as well as a majority of the quantum spin models, may not be
solved analytically apart from a few special cases. Moreover, unlike in classical problems, numerical
calculations are generally not helpful. The exponentially large size
of the Hilbert space prevents an exact diagonalization of $H$ and
the fermion sign problem renders Monte Carlo calculations
inefficient. Since the fermion wavefunction changes sign under interchange of
particles, calculating the partition function requires summing a large
number of positive and negative terms, which results in large numerical
error. Therefore to solve the problem one must try to deduce behaviours that
might appear based on toy models and/or use basic physical notions such as
{\it universality}. In classical systems, universal, system independent physics
emerges when the correlation length becomes much larger than the interatomic
spacing. The difference in the order of the magnitudes of the two
characteristic lengths is responsible for a length scale separation and
leads to system independent behaviour. In quantum system, the scale
separation necessary for the existence of universal phenomena, is provided
by the large characteristic energy scale of electrons in solids ($t$
or $U$ in the Hubbard model), which is orders of magnitude larger than
room temperature.  

\subsection{Ordering in a Mott insulator}

Let us now focus on the behaviour at zero temperature of Mott insulators,
which can be described by a spin Hamiltonian such as the Heisenberg model
Eq.~(\ref{HeisHam0}) and its variants. 

\begin{figure}
\centerline{\includegraphics[height=3cm,width=4.5cm]{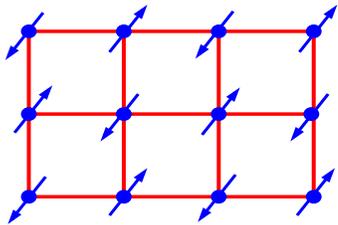}}
\caption{The antiferromagnetic (or N\'eel) state on a square lattice.}
\label{fig:neel}
\end{figure}

\begin{figure}
\centerline{\includegraphics[height=5cm,width=6.5cm]{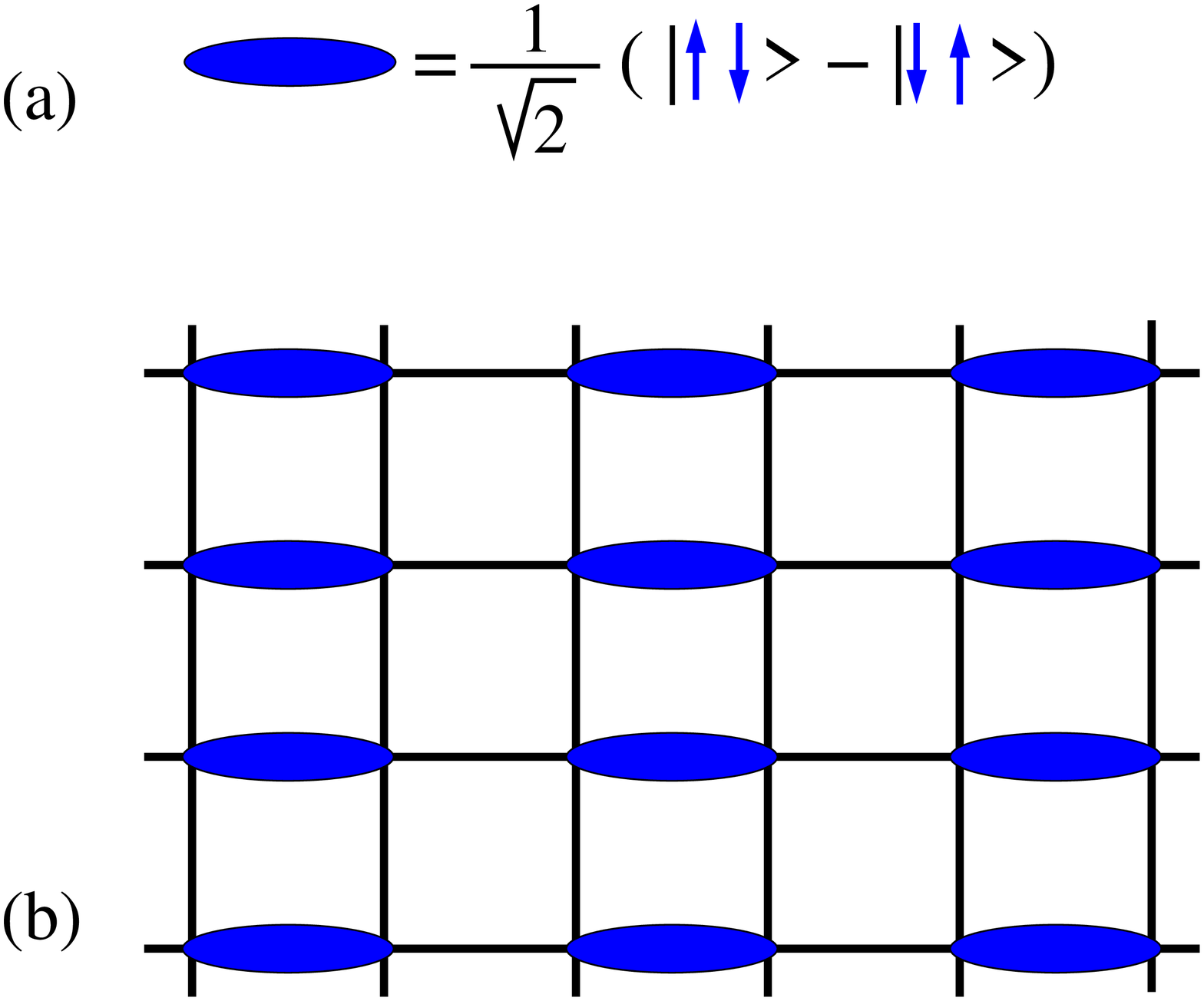}}
\caption{(a) A valence bond (or a dimer) is a singlet pair formed by two neighbouring
  spins on a lattice. (b) The Valence Bond Solid (VBS) formed by crystalline
  ordering of these dimers.}
\label{fig:vbs}
\end{figure}

At $T=0$, Mott insulating systems usually break symmetries and develop
order. For exemple, the $2d$ Heisenberg model on the square lattice develops antiferromagnetic
(N\'eel) long range order associated with the breaking of spin rotation
symmetry (see Fig.~\ref{fig:neel}). Another possibility is that two neighbouring spins
prefer to first pair themselves in a singlet $\frac{1}{\sqrt{2}}(\ket{\uparrow
  \downarrow}-\ket{\downarrow \uparrow})$, forming a valence bond (a
``dimer''). This is denoted in Fig.~\ref{fig:vbs}A by an oval. These dimers can then order amongst themselves, forming a
valence bond solid (see Fig.~\ref{fig:vbs}B) which breaks translational symmetry. In
both cases, these phases can be described by a local order parameter.
Transitions in or out of these phases are then formulated within Landau's
traditional theory of phase transitions, which expands the free
energy in powers of this order parameter. 

\begin{figure*}
\centerline{\includegraphics[height=7cm,width=15cm]{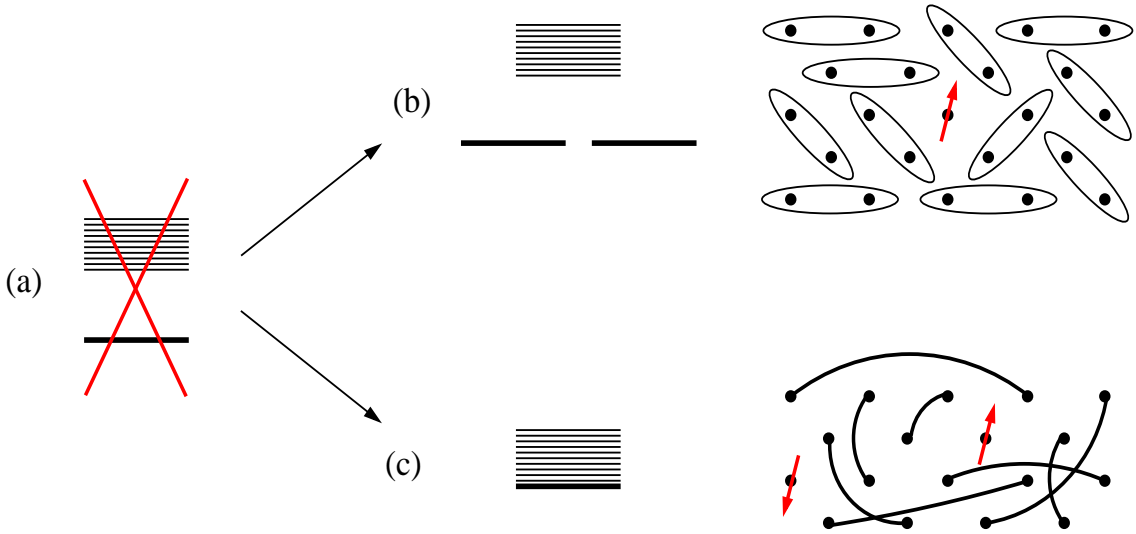}}
\caption{Possible energy spectra (and associated real-space cartoon picture)
  of a spin liquid state with a disordered ground state in the thermodynamic
  limit. (a) For $S=1/2$
  spin systems with odd number of spins per unit cell, the Hastings' theorem forbids
  non-degenerate ground state with a finite energy gap. (b)
  Topological spin liquids possess a multiply-degenerate ground-state (with a
  topology-dependent degeneracy) with a gap to all excitations in the
  bulk. The cartoon picture presents a RVB liquid state of dimers with a
  delocalized free spinon $S=1/2$ excitation. See Sec.~\ref{sec:topo}. (c)
  Critical spin liquids possess a ground-state with power-law correlations
  and gapless excitations. In real space, critical spin liquids can
  typically be seen as valence bonds on many length scales. The excitations
  are spinons interacting with this valence-bond background. See Sec.~\ref{sec:crit}.}
\label{fig:hastings}
\end{figure*}

Our interest in the following sections lies elsewhere. We want to consider
$S=1/2$ systems which do not break any symmetries when cooled to zero
temperature - they do not possess long-range order. These systems, dubbed spin liquids, possess however some kind of
exotic order, which cannot be understood in terms of a local order parameter.
A theorem recently proven by Hastings \cite{Hastings} guarantees the
existence of exotic quantum ground states in Mott insulators with no broken
symmetries. This theorem, a generalization of the
Lieb-Schultz-Mattis theorem~\cite{LSM} to dimensions larger than $1$, states that for $S=1/2$ 
systems with one spin per unit cell on a two-dimensional lattice with periodic boundary
conditions (on a torus), in the absence of symmetry breaking the
ground state is separated from the first excited state by an energy gap that vanishes in the
thermodynamic limit:   $E_1-E_0<\frac{\ln(L)}{L}$, for a system of size
$L$ by $L$. This means that in the thermodynamic limit, there
cannot be a singly-degenerate quantum paramagnet with a finite energy gap (see Fig.~\ref{fig:hastings}a). In the following sections, we will
describe two ways out of this situation and the exotic properties of the
associated systems. Firstly we discuss in Sec.~\ref{sec:topo} topological spin liquids, where the disordered ground
state is degenerated (see Fig.~\ref{fig:hastings}b) according to the topology of the
system. These systems have an emergent gauge structure and admit excitations
with fractional quantum numbers. In Sec.~\ref{sec:crit}, we will describe critical or algebraic spin liquids,
with gapless excitations above a ground-state that possesses power-law
correlations (see Fig.~\ref{fig:hastings}c). Such systems behave without
fine-tuning as if they were at a critical point.

\section{Topological spin liquids}
\label{sec:topo}

\subsection{Description of topological spin liquids}

As their name suggests, topologically ordered spin liquids have peculiar behaviour
associated with the topology of the system. They have a degenerate
ground state in the thermodynamic limit with a degeneracy that depends on the
topology of the system. The ground state wavefunctions, associated with the degenerate states, cannot be
distinguished by any local measurement but are globally distinct. These
states also have an emergent gauge structure which supports gapped
particle-like excitations, with quantum numbers that cannot be built up from
electron quantum numbers. It is possible to partially classify such spin
liquid phases in terms of the symmetry group of the emergent gauge theory. General discussions on topological order and its classification can be found in the book of
Wen~\cite{BookWen}. For simplicity, we will focus here on the simplest
example of a topologically ordered system: a spin liquid with $Z_2$ topological order on a $2D$ square lattice.

\begin{figure}
\centerline{\includegraphics[height=6cm,width=7cm]{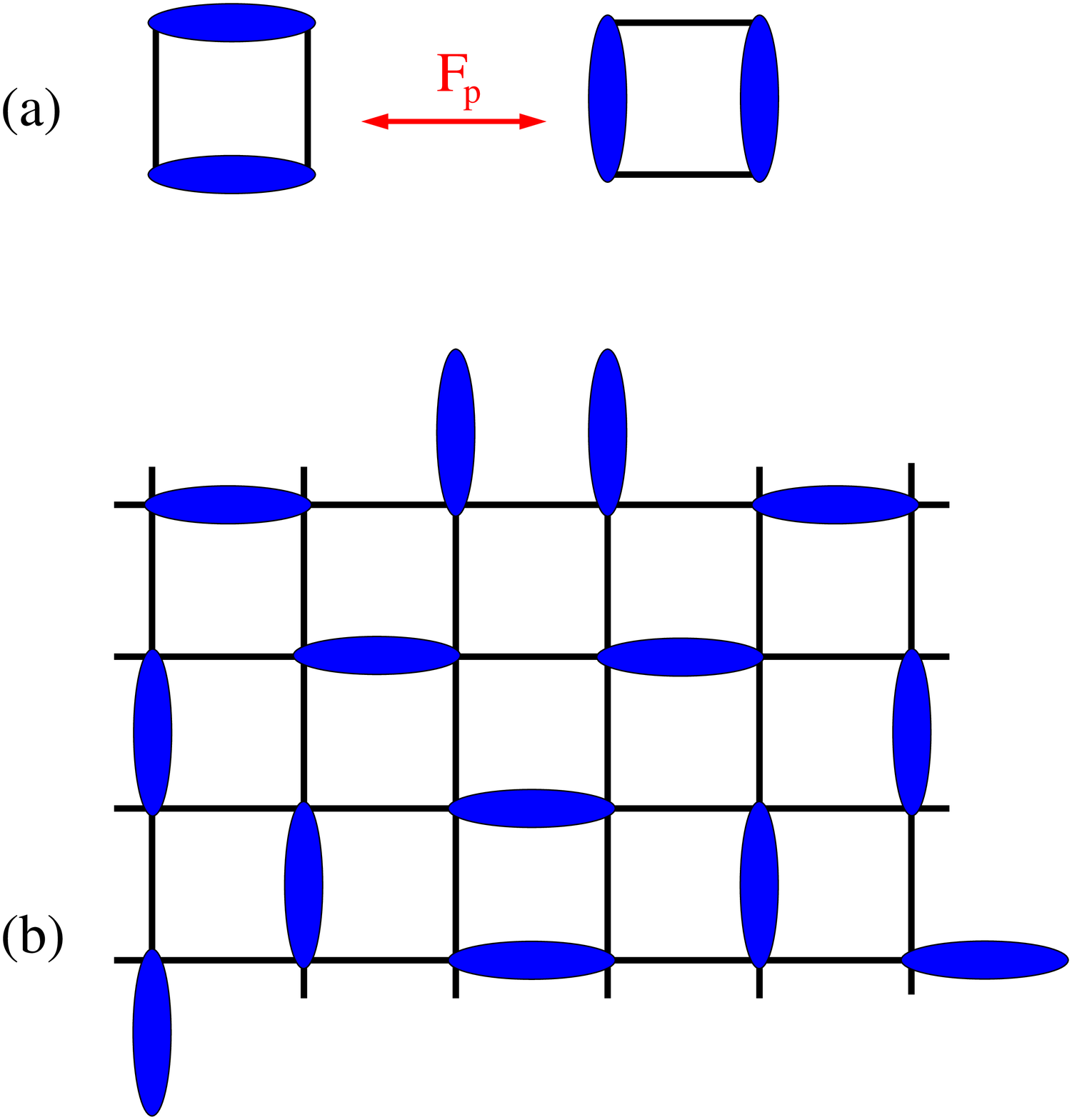}}
\caption{(a) The two resonating states connected by quantum fluctuations on
  a plaquette. The flux operator $\it{F}_p$ flips the bonds between the two
  resonating states (see Sec.~\ref{sec:gauge}). (b) A Resonating Valence Bond
  (RVB) liquid is a disordered state of valence bonds.}
\label{fig:rvb}
\end{figure}

As proposed by P.W. Anderson~\cite{anderson}, spin liquids can appear as a result
of melting a valence bond solid. Consider a generalized Heisenberg model
with not only nearest neighbours interactions 
\beq
H=\sum_{ij}J_{ij}{\bf S_i}\cdot {\bf S_j}
\label{HeisHam}
\eeq
with $J_{ij}>0$. As already mentioned, in some cases, two neighbouring
spins will form, due to the antiferromagnetic interaction, a singlet pair
(a valence bond). These singlets can possibly align to form the valence bond
solid (VBS), depicted in Fig.~\ref{fig:vbs}b. The bonds between four neighbouring sites can also
fluctuate quantum mechanically between two possible positions, as shown in
Fig.~\ref{fig:rvb}a. These resonances between the bonds can lead to the
melting of the solid to another state of matter with no long-range
correlations of any kind: a resonating valence bond (RVB) liquid (Fig.~\ref{fig:rvb}b). The RVB liquid can be described using a wavefunction
which is a superposition of many valence bond configurations.

\begin{figure}
\centerline{\includegraphics[height=8cm,width=7cm]{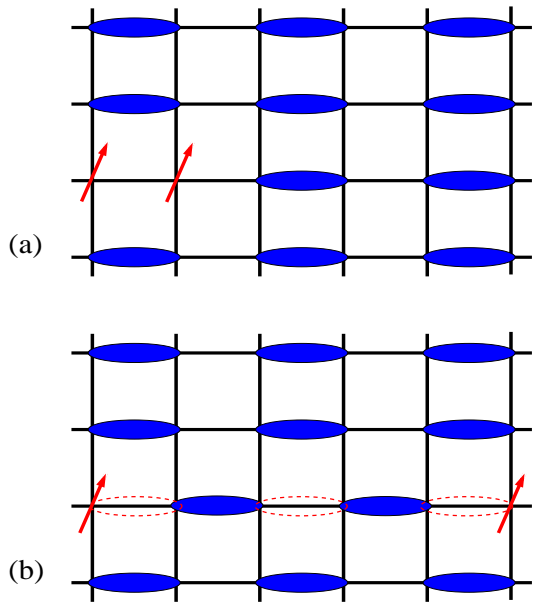}}
\caption{(a) Breaking of a valence bond results in two spinon excitations.
  (b) In a VBS, two separated spinons have created a mismatch in the
  arrangement of dimers: the energy cost will be linear in the distance
  between spinons.}
\label{fig:spinon}
\end{figure}

What are the excitations of these two different phases? A valence bond may be broken, at an
energy cost of the order of $J$, to create an $S=1$ (gapped) excitation in which the
spins are aligned (see Fig.~\ref{fig:spinon}a). Let's try to separate these two $S=1/2$ ``spinons''. The
energy cost for this separation will grow {\it linearly} with separation in
the VBS, because there will be a mismatch in dimer aligment
all along the path separating the two spinons (see Fig.~\ref{fig:spinon}b). The spinons are therefore
{\it confined} in the VBS, similarly to quarks in elementary particles. They cannot exist as finite
energy excitations. However in the RVB liquid state, the region between the
two spinons is not a defect line, as in the solid, therefore the energy
costs stays finite with separation. The reorganization of the valence bonds
to accomodate the extra spinon is local - or at least the energy cost is local. Therefore the spinons are {\it deconfined} in the liquid state (see the cartoon picture of Fig.~\ref{fig:hastings}b). Furthermore the
spinons carry the $S=1/2$ spin of the electron, but do {\it not} carry any
additional electrical charge. The charge density in the topological spin liquid is
uniform and equal to one electron per cell. The existence of these
fundamental excitations which cannot be described by electron quantum
numbers is referred to spin-charge separation. Quantum numbers
are said to be fractionalized, since the spin $S=1/2$ of the spinon is half the
one of the conventional spin-flip excitation with $S=1$.

\subsection{Gauge theory formulation}
\label{sec:gauge}

Due to their physical origin, the spin interactions in the Hamiltonian~(\ref{HeisHam}) are typically local.
There is no explicit long-range interaction. How can we then understand the
confinement/deconfinement properties of the spinons, which might be far apart? The natural
framework for this is {\it gauge theory}. Let us now formulate the quantum
dynamics of the valence bonds in such terms. 

To formulate the problem in terms of a gauge theory, consider a ``spin'', on each lattice link, between two sites $i$ and $j$. This ``spin'' is
described by a set of Pauli matrices ${\bf \sigma}_{ij}$, the $x$ component
of which takes on the value $\sigma^x_{ij}=1$ if there is no valence bond
(no dimer) on the given link $(i,j)$ and $\sigma^x_{ij}=-1$ if there is one.
We want to define an operator which, when acting on four spins on a
plaquette, flips the bonds between the two resonating states, as depicted in
Fig.~\ref{fig:rvb}a. This is easily done using operators that anticommute with each other, such as Pauli matrices. The bond flipping operator, called the plaquette flux operator,
is defined on the four sites $i,j,k,l$ around the plaquette as
$\it{F}_p=\sigma^z_{ij}\sigma^z_{jk}\sigma^z_{kl}\sigma^z_{li}$. The
Hamiltonian which describes the resonating valence bonds is a sum over all
plaquettes of the flux operators $H=-K\sum_{p(ijkl)} \it{F}_p$. 
Up to this point we have not taken into account the
constraint that each spin can form only one singlet, {\it i.e.} only one
valence bond comes from each site. Such a local constraint on the Hilbert
space of this ``quantum dimer model'' may be encoded in a {\it gauge charge}
$Q_i$ for each site $i$. The gauge charges are defined as the product of the bonds at
a given site and we constrain them to be equal to $-1$, $Q_i=\prod_{j=1}^4
\sigma_{ij}^x=-1$. This constraint is still not sufficient, as the number of
bonds at each site may still be equal to one or three. To get rid of the
possibility of three bonds, we add a term to the Hamiltonian, which
minimizes the number of bonds at each site $-J\sigma^x_{ij}$, where $-J$ is
the large energetic gain of having no dimer on a given link ($J>0$). Therefore the
full formulation of the problem is in terms of the Hamiltonian:
\beq
H=-K\sum_{p(ijkl)} \sigma^z_{ij}\sigma^z_{jk}\sigma^z_{kl}\sigma^z_{li}-J\sigma^x_{ij}
\label{GaugeHam}
\eeq   
with the constraint on the gauge charges $Q_i=\prod_{j=1}^4
\sigma^x_{ij}=-1$. This is equivalent to the formulation of a $Z_2$ gauge theory~\cite{Kogut}.
Note the, rather vague at this stage, analogy with the electrodynamics Hamiltonian
$H=\textbf{B}^2+\textbf{E}^2$, supplemented by ${\bf \nabla} \cdot
{\bf E}=0$.

Since the gauge charge operator and the Hamiltonian commute $[Q_i,H]=0$, one
can simultaneously diagonalize $Q_i$ and $H$. However not all eigenstates
$H\ket{E}=E\ket{E}$ of the Hamiltonian are physically relevant states, as
they do not satisfy the odd-bond-per-site constraint. The projection
operator $\it{P}=\prod_i \frac{1}{2}(1-Q_i)$ constructs a state $\tilde{E}$, that
obeys the bond constraint, out of the eigenstates $E$ of the Hamiltonian. It
is also worth noting that the Hamiltonian is invariant under a gauge
transformation $\sigma^z_{ij}\rightarrow \epsilon_i \sigma^z_{ij}
\epsilon_j$ with $\epsilon_i=\pm 1$. Note the analogy with the gauge
transformation ${\bf A} \rightarrow {\bf A}+{\bf\nabla } \phi$ in electromagnetism. The
physical observables, the bond field $\sigma^x_{ij}$ (the ``electric field''
in the analogy) and the plaquette flip $\it{F}_p$ (``magnetic flux'') 
are also gauge invariant. However the gauge
field itself is not invariant under the gauge transformation. We ought to
speak in our case of a gauge redudancy rather than gauge invariance,
since the states related by a gauge transformation are physically
equivalent. The total Hilbert space for a $N$ site square lattice model with
$2N$ links has $2^{2N}$ states. Each gauge inequivalent class, however, has a
redudancy of $2^N$. Therefore the number of physical distinct states is
$\frac{2^{2N}}{2^N}=2^N$, which correspond to the fluxes through the $N$
plaquettes $\it{F}_p=\pm 1$.

\begin{figure}
\centerline{\includegraphics[height=0.8cm,width=7cm]{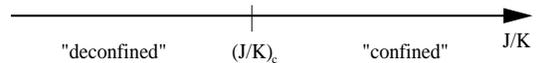}}
\caption{Phase diagram of the $Z_2$ gauge theory model
  Eq.~(\ref{GaugeHam}). For small values of $J$, the plaquette flux is fixed and the spinons are deconfined in the spin liquid phase. For $J \gg K$, the bond field is fixed, which corresponds to confined spinons in the VBS state.}
\label{fig:topphd}
\end{figure}

We can now gain insight about the confined VBS and
deconfined RVB liquid phases of the dimer model from the known phase diagram
of a $Z_2$ gauge theory, which is depicted in Fig.~\ref{fig:topphd} in terms
of the parameters $J$ and $K$. In the $J=0$ limit, the plaquette flux is
fixed $\it{F}_p\approx 1$, as such a flux configuration minimizes the first
term in the Hamiltonian Eq.(\ref{GaugeHam}). The conjugate bond field
$\sigma^x_{ij}$, however, is strongly fluctuating. This corresponds to the resonating
valence bonds in the deconfined spin liquid state. In the opposite limit,
$J \gg K$, the bond field is fixed and the plaquette flux fluctuates, which
describes the behaviour of the VBS. The dimer model was
formulated assuming the energy cost $J$ of having a bond at a given link is
large compared to $K$, to fulfill the one-bond-per-site constraint. From the
phase diagram in Fig.~\ref{fig:topphd}, we see that for the quantum dimer
model on a square lattice, this set of parameters corresponds to the
{\it confined state}. In order for the system to be in the deconfined phase, one would have to relax this constraint on the value of $J$ or
the gauge charge constraint.  The latter scenario is realized, for example, for
{\it non-bipartite lattices}, such as the triangular lattice for which a RVB
liquid phase exists~\cite{qdm}.

Many things are known about the $Z_2$ gauge theory, which is dual to the $2D$ Ising model in a
transverse field. In particular, we have a way to characterize the different
phases through the {\it Wilson loop operator} $W_L$, defined as the product of
the gauge operators $\sigma^z$ around a loop of size $L$: $W_L=\prod_{i
  \in L} \sigma^z_i$. The expectation value of the operator for a loop of
size $L$ decays as an exponential of the perimeter of the loop
$<W_L>=e^{-cL}$ in the deconfined phase and as the exponential of the area
substanded by the loop $<W_L>=e^{-cL^2}$ in the confined phase. We refer to
Ref.~\cite{Kogut} for an intuitive derivation of these two different behaviours.

\subsection{Analogy with electromagnetism, vison excitations and ground-state
degeneracy} 

\begin{figure*}
\centerline{\includegraphics[height=4.5cm,width=11cm]{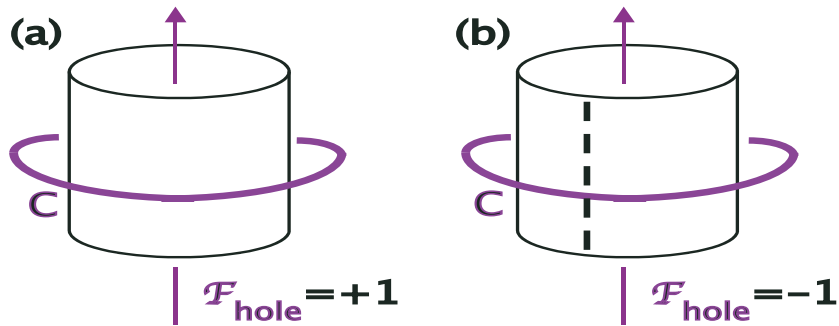}}
\caption{The two degenerate states in the deconfined (RVB) phase on a
  cylindrical topology. (a) No vison is
  present. (b) A vison has threaded the cylinder. All the links that cross
  the dashed line have $\sigma_i^z=-1$ (whereas $\sigma_i^z=1$ for the rest
  of the system).}
\label{fig:cylinder}
\end{figure*}

The vaguely sketched analogy between the $Z_2$ gauge theory of spin liquids
and electrodynamics may be further exploited. It can easily be shown, that the bond field and the
plaquette flux do not commute whenever the bond surrounds the plaquette. Based on the Hamiltonian
Eq.(\ref{GaugeHam}), we can identify the first with the electric field and the
latter with the magnetic flux. Being non-commuting fields, the Heisenberg
uncertainty principle applies and one field has to fluctuate
if the other one is fixed. 

In the deconfined spin liquid phase, the flux through all 
the plaquettes in the ground state is constant ($\it{F}_p\approx 1$) and the
bond field fluctuates a lot (the dimers are disordered), as
argued above. If the flux through one plaquette is $-1$, an excitation is
created in the liquid, which has an energy cost of the order of $K$. This
gapped excitation is called a {\it vison} and takes its name from an Ising
vortex. Two visons on the same plaquette ``annihilate'' since the flux through the plaquette
can only be $\pm 1$ - hence the reference to Ising. The vison is a non-magnetic
($S=0$) excitation, of a different kind than the spinons described earlier. We are describing a liquid
phase and both visons and spinons are finite energy excitations. How can we reconcile this with the
Hastings theorem about the existence of low energy excitations, which tend to
zero in the thermodynamic limit? The existence of the visons 
allows us to understand how this theorem may be satisfied. This
theorem is fulfilled because the ground state is {\it topologically}
degenerate, when the model is defined on nonsimply connected spaces
such as a torus.
In conventional paramagnets, the properties such as
ground-state degeneracy are not affected in the thermodynamic limit by the
choice of boundary conditions. This is not true anymore in a topological
spin liquid. 

To see this, let us take the cylinder topology, where a $2d$ system has periodic boundary conditions
in one direction and free in the other.  In the deconfined phase, locally the
flux through each plaquette is $\it{F}_p\approx 1$, but what about the hole
in the cylinder? Consider the flux through any curve $C$, that encircles
the cylinder $\it{F}_{hole}=\Pi_{i\in C} \sigma_i^z$
(Fig.~\ref{fig:cylinder}). In the deconfined phase of the gauge theory, we
can have both situations where there is a flux through the hole, or no flux
through it. In the limit $K\rightarrow \infty$, a simple ground-state is obtained when all the link fields are
positive $\sigma_i^z=1$. There is no vison present and $\it{F}_{hole}=1$
(Fig.~\ref{fig:cylinder} (a)). Now if we flip to $\sigma_i^z=-1$ all the
dimers along a column of horizontal bonds that runs the length of the
cylinder, we have a new state where $\it{F}_{hole}= -1$ and where a vison
has ``threaded the hole of the cylinder'' (Fig.~\ref{fig:cylinder} (b)). On
a {\it finite} system, this is an excited state when $J \ne 0$, but in the thermodynamic
limit the energy cost of creating this line of defect tends to zero and this
state also becomes a ground state. Hence the two-fold degeneracy of the ground state. 
The multiplicity of the degeneracy of the ground state therefore depends on the {\it topology} of the space of system; more
specifically on its genus in our example. For example, the degeneracy
is four-fold on a torus. Ref.~\cite{vison} describes in great detail the
possible interpretations of this degeneracy and its topological nature. 

Moreover, from this example we see the hidden topological order is a global property of the whole
system and cannot be seen on the local scale ($\it{F}_p= 1$ everywhere in
the ``bulk'' for both degenerate ground states). It is easy to guess that it is impossible to deduce
whether a vison has threaded or not the hole of the torus by looking locally
at the dimers. Therefore one cannot distinguish between the different ground
state wave functions by any local probe or order parameter; they are
topologically distinct.

In the VBS phase, the bond (electric)
field is fixed (the dimers order) and the magnetic flux fluctuates. This
means that visons have proliferated, they are {\it condensed} in the VBS. 
A site with a spinon has no connecting valence bonds and since
the gauge charges are defined in terms of the number of bonds in the system,
such a spinon carries a gauge charge $Q_i=-1$, which can be thought of as an ``electric'' charge. 
These electric charges cannot propagate through the fluctuating ``magnetic''
fluxes. This is another way to see that the spinons are confined and no
longer present as finite excitations in the valence bond solid phase. 

With this electromagnetic analogy, we can also understand the nature of the 
interaction between the different excitations. A vison can be though of as
a ``magnetic'' flux. Hence if a vison is taken around a spinon (the electric
charge), analogously to the Aharonov-Bohm effect, it will result in a sign change in
the wavefunction of the spinon $\Psi_s\rightarrow -\Psi_s$. This may be
thought of as a long-range ``statistical'' interaction mediated by the gauge
field.

\subsection{Towards realistic models}

We have shown how by melting a VBS into a RVB liquid one obtains deconfined
finite energy excitations - spinons. These can be described in terms of a
$Z_2$ gauge theory. Exploiting the analogy with electromagnetism, one can
identify another type of excitation - visons, which correspond to magnetic
fluxes. By considering the topology of the space, on which the model is defined, one discovers the
ground states are degenerate. Hence Hastings' theorem is satisfied, and
exotic quantum ground states exist. Having identified the finite energy
excitations, one can complete the electromagnetic analogy by considering the
Aharonov-Bohm type interactions between the  ``electric" and ``magnetic" charges.

Here we have only focused on phenomenological description of spin systems
which can have a topological $Z_2$ spin liquid phase. Actually, there are
already several models which can be shown (sometimes on
rigorous grounds) to sustain $Z_2$ topological order. Without being exhaustive, one
can mention quantum dimer models on non-bipartite
lattices~\cite{qdm}, Ising-like models with
multiple spin interactions~\cite{Kitaev}, rotor bosonic
models~\cite{Motrunich} and more recently models Heisenberg-like spin models with original
$SU(2)$ symmetry~\cite{SU2} or without~\cite{kagomeising}. We should also
note that the $Z_2$ spin liquid is only the tip of the iceberg: there are
many much more intricate topologically ordered phases possible, some with
fractional and even non-Abelian statistics~\cite{kitaev2}. These phases are very promising
candidates as possible realizations of quantum computers, as they could be
topologically protected from decoherence effets~\cite{ioffe}. We also have
kept the discussion to $2d$ systems, but this is not essential to obtain
topological order.  It is, however, necessary to obtain the fractional
statistics and non-abelian properties.

Finally, we note that at this point theoretical models are sadly not very
realistic. It remains a huge challenge to develop theoretical techniques to look for topological spin liquids in realistic
models and find them in
the laboratory. There is no experimental
evidence, except for the fractional quantum Hall effect, as to the existence
of topologically ordered phases. It was theoretically proposed~\cite{SenthilFisher} that
topological order and visons could exist in high $T_c$ cuprates superconductors.
Later experiments~\cite{YBaCuO} disproved this statement.
Nevertheless, these ideas and the proposed experimental set-up to detect visons
could be useful to eventually experimentally engineer a real topological
ordered phase, for example in a Josephson Junction array. The key
ingredient is certainly to try to disorder the system as much as possible.

\section{Algebraic or Critical Spin Liquids}
\label{sec:crit}

\subsection{Description of algebraic spin liquids}

Besides topological degeneracy, an alternate way in which an apparently featureless quantum
paramagnet can remain ``gapless" as required by the  Hastings' theorem,
is by having gapless {\it bulk} excitations - {\it i.e.} being a ``critical"
quantum phase.
Such critical quantum ground states 
have power-law forms for correlation
functions, but no true long-range order. The lack of a finite order parameter 
is necessary for a phase to be called a spin liquid. In some sense, these
systems behave as if they were at a critical point without being tuned.
A physical picture in terms of the
spin degrees of freedom for such critical spin liquid phases can be visualized in terms
of valence bonds between pairs of spins separated by arbitrary distances.
The valence bond lengths follow a power-law
distribution. These valence bonds quantum-mechanically fluctuate,
which produces the continuum of excitations mentioned above. The spinons (unpaired
spins), unlike in topological spin-liquids, are strongly interacting with
this valence bond background and cannot be considered as free particles.
A cartoon picture of a critical spin
liquid is given in Fig.~\ref{fig:hastings}c.

Much less is known about critical spin liquids than topological
spin liquids, as introduced in Sec.~\ref{sec:topo}. Technically, the theoretical models involved are more complicated and even toy models are
difficult to construct. Apart from the general considerations given in the
preceding paragraph, there is no clear overall theoretical picture of all the
generic features of critical spin liquids, although some progress has
recently been made in this direction~\cite{Hermele}. However, an advantage over
topological spin liquids is that there are several experimental spin systems which are
good candidates for being a critical spin liquid,
and can thus be confronted with theory.
We
will describe one particular theoretical scenario which gives rise to a critical
spin-liquid phase, which is consistent with current experiments on one promising compound.
This example will let us identify some of the
elements that are needed to form a critical spin-liquid. 

\subsection{Triangular lattice $S=1/2$ antiferromagnets - Experimental
  evidence of spin liquid behaviour}

As already mentioned, the key idea to obtain a spin liquid is to enhance the
conditions that can supress magnetic order, namely using low spins ($S=1/2$
or $S=1$), low dimensional systems, and
geometric frustration, as in non bipartite-lattices~\cite{misguich}. The 1d Heisenberg chain is the simplest
example of a critical system - it is much harder to find one in 2d. Our interest lies in 2d systems, thus we will
consider a spin $1/2$ antiferromagnetic sytems on the
triangular lattice. Numerical calculations have determined that the ground
state of the isotropic Heisenberg model on the triangular lattice has long
range coplanar magnetic order, with spins ordering in a 120 degrees fashion. This order is the same
as for the classical ground state of the model, which results from the minimization of the
frustrated interactions. However, in the $S=1/2$ quantum system, the magnetization is
significantly reduced compared to the $S \rightarrow \infty$ classical limit, due to quantum
fluctuations. This numerical result, along with recent experiments, gives one reason to
believe that a spin liquid state is perhaps accessible in such triangular lattice antiferromagnets.

\begin{figure*}
$\begin{array}{lr}
\includegraphics[height=5.5cm]{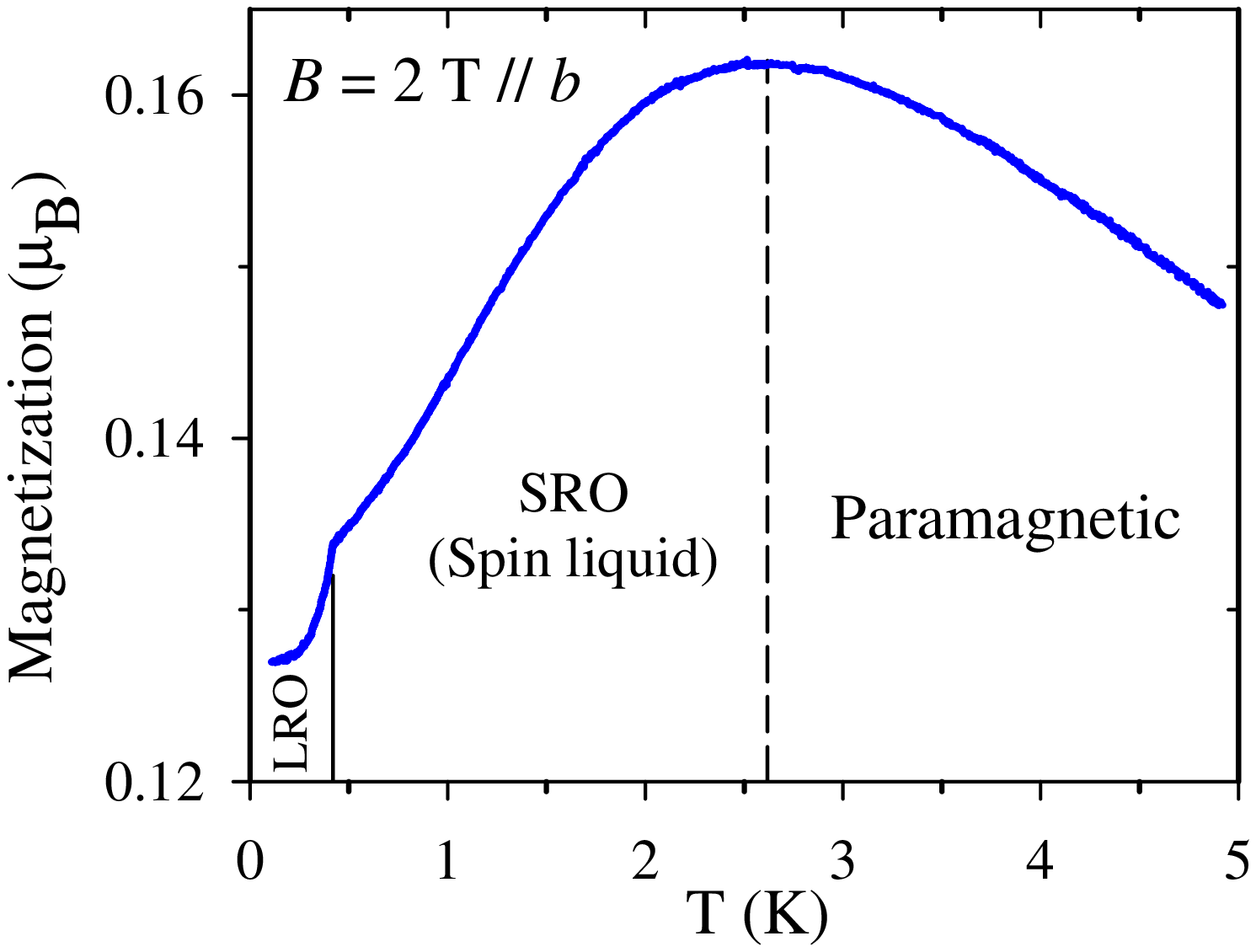} \hspace{0.9cm} & \includegraphics[height=5.5cm]{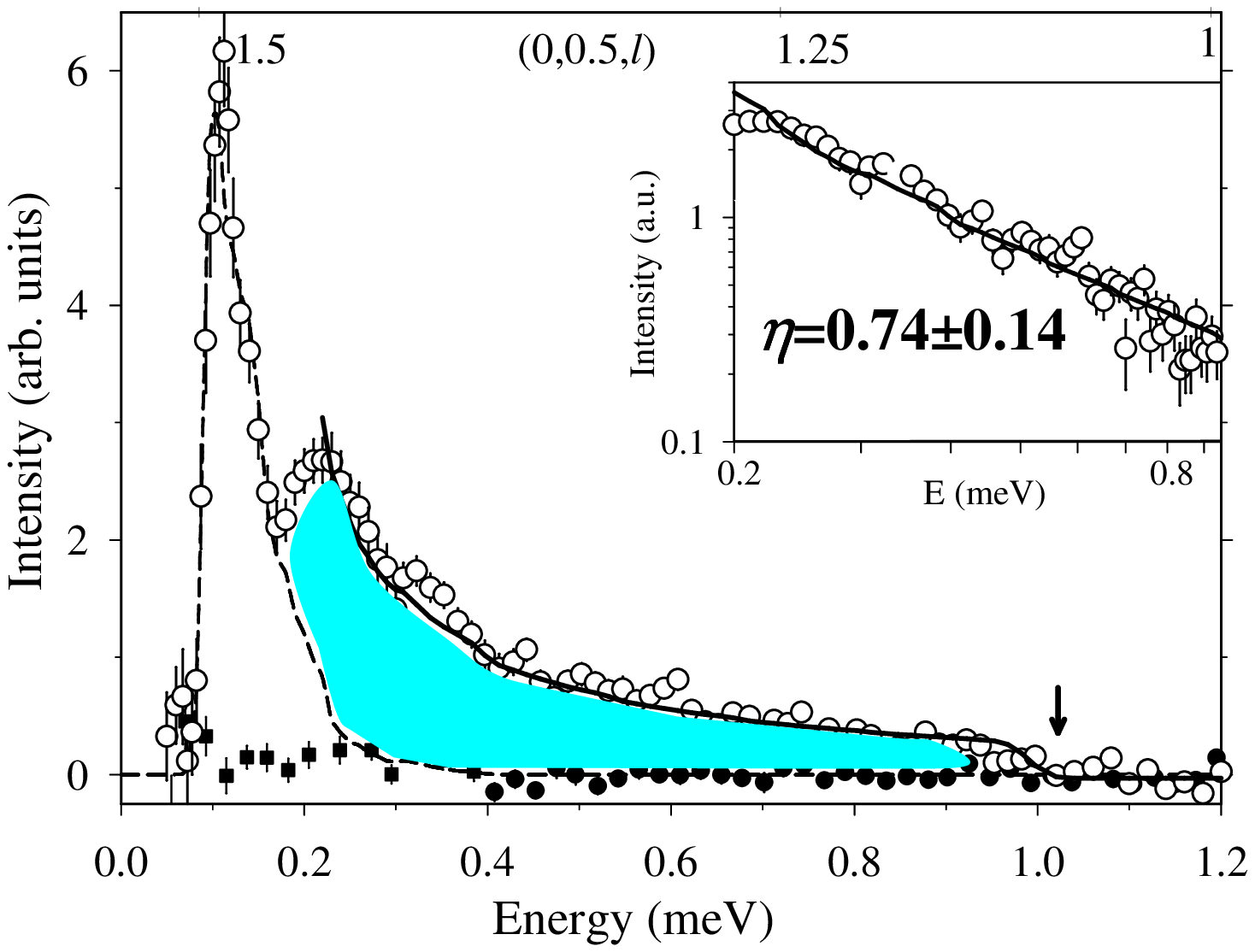}
\end{array}$
\caption{Experimental results obtained on Cs$_2$CuCl$_4$. Left: Low field
  magnetization versus temperature (taken from Ref.~\cite{coldea}d). 
  At the transition from the long-range ordered (LRO) state at low temperatures to
  the short-range ordered (SRO) spin liquid at intermediate temperatures
  the magnetization exhibits a cusp. The broad peak at the transition from the spin liquid to
  the paramagnet present at high temperatures is characteristic of
  short-range antiferromagnetic correlations. Right: Neutron scattering
  intensity as a function of energy transfer (taken from
  Ref.~\cite{coldea}b). The sharp peak at low energy is the signature of
  long-range ordering. The broad continuum at high energies can be fitted by
  a power-law (see inset) and is ascribed to the low-energy excitations of a
  critical spin liquid state.}
\label{fig:expPD}
\end{figure*}

\begin{figure}
\centerline{\includegraphics[height=5cm,width=5cm]{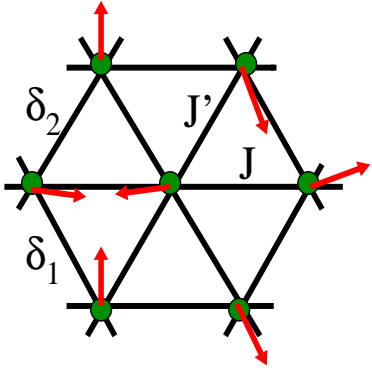}}
\caption{The long-range coplanar ordered state adopted by the $S=1/2$
  antiferromagnet Cs$_2$CuCl$_4$ (see Ref.~\cite{coldea}). The bond directions  $\delta_1$ and $\delta_2$ and the
  interaction strengths $J$ and $J'$ in the effective Hamiltonians Eq.(\ref{MeasCritHam1}) and
  Eq.(\ref{MeasCritHam2}) are also shown.}
\label{fig:critlat}
\end{figure}

Currently, two experimental systems are likely candidates for spin
liquid behaviour in the triangular $S=1/2$ antiferromagnet: an organic
material $\kappa$-(ET)$_2$Cu$_2$(CN)$_3$ (Ref.~\cite{kappa}) and a transition metal
material Cs$_2$CuCl$_4$ (Ref.~\cite{coldea}). We will focus the discusion on the
latter system for which the Cu atoms have nine electrons on the $d$ shells and the one
missing electron results in the $S=1/2$ moment. The experimental phase
diagram for Cs$_2$CuCl$_4$ (Fig.~\ref{fig:expPD}), shows a long range magnetically
ordered phase at very low temperatures and a ``spin liquid" in an intervening
temperature regime. At higher temperatures the spins on the Cu atoms fluctuate
independently and the magnetic susceptibility obeys the traditional Curie-Weiss
law. The planar triangular lattices in the Cs$_2$CuCl$_4$ crystal are stacked
on top of one another, with a large interplanar distance. Therefore the
in-plane interactions $J$ and $J'$, as depicted in Fig.~\ref{fig:critlat}, are
dominant. The magnetic long range ordered state is coplanar as predicted by
numerical studies. Upon moving through the lattice the spins spiral in the plane.
However, since the exchange interactions $J$ and $J'$ differ in magnitude the
wavelength of the spiraling spins is incommensurate with the lattice,
resulting in the spin ordering represented in Fig.~\ref{fig:critlat}.

In Cs$_2$CuCl$_4$ the experimentally detected ``spin liquid"  only exists at finite temperature, whereas a theoretical spin liquid is well defined only
at $T=0$.  Despite this drawback, there are two significant advantages of Cs$_2$CuCl$_4$ that compensate.
Firstly, it is possible to grow large high purity single crystals, which may be used in neutron
scattering. The main quantity which can be accessed in neutron scattering experiments is
the magnetic structure factor, which is the Fourier transform of the
spin-spin correlation function. At low temperatures, the experiments show magnetic Bragg peaks at wavevectors
corresponding to the wavelength of magnetic ordering, which detect the
spiral long-range ordered state (see Fig.~\ref{fig:expPD}).  Inelastic neutron scattering
experiments at higher temperatures, where the long range order is destroyed,
can measure the strength of quantum fluctuations
in the putative ``spin liquid".  Strikingly, these measurements show a
power law dependence on energy transfer $\omega$ of the structure factor 
\beq
\left< S^+({\bf q},\omega) S^-({\bf -q},-\omega) \right> \sim
\omega^{\eta-2}
\eeq 
with $\eta\sim 0.75$ for wavevectors ${\bf q}$ chosen close to those of
the Bragg peaks. 
This power-law feature is 
suggestive of a critical spin-liquid phase.
The intensity of the scattered beam as a function of
$\omega$ is plotted in Fig.~\ref{fig:expPD}b. Two features can be observed.
Firstly, a sharp low-energy peak which shows up at low temperature is the
signal of spin-waves associated with the long-range order. This peak
disappears for $T>T_c$. Secondly, a broad continuum scattering which extends
up to high $\omega$ suggests many low-energy excitations. The number of
these low-energy excitations increases as the  aforementioned power law, as the transfer energy is lowered (see
log-log scale in the inset of the figure). 

The second advantage of Cs$_2$CuCl$_4$ is that the exchange interactions $J$
and $J'$ are very small, of the order of a few Kelvins. Therefore one can experimentally measure the
parameters of the effective Hamiltonian. These experiments are conducted in
high magnetic fields, where the spins are forced to be polarized along the
external field. This is possible because the Zeeman energies involved are of
the order of a few Kelvins, corresponding to magnetic fields of a few Teslas,
which can easily be reached in the laboratory. Using a
neutron beam a single spin may be flipped and by measuring the energy-momentum
dispersion relation for this flipped spin, the values of the exchange interaction may
be extracted. The Hamiltonian for a single layer has the form of nearest
neighbour exchange interactions on the triangular lattice (see Fig.~\ref{fig:critlat})
\begin{eqnarray}
H_0 & = & \sum_{{\bf r}}[J{\bf S}_{{\bf r}} \cdot {\bf S}_{{\bf
      r}+\delta_1+\delta_2} \nonumber \\
&~ & ~~~~~+J'({\bf S}_{{\bf r}} \cdot {\bf S}_{{\bf r}+\delta_1}+{\bf S}_{{\bf r}} \cdot {\bf S}_{{\bf r}+\delta_2})].
\label{MeasCritHam1}
\end{eqnarray}

In addition there is a term due to spin-orbit coupling, called the
Dzyaloshinskii-Moriya interaction. In most systems this term is small, but  it
may not be neglected here, since the exchange interactions are small. The full Hamiltonian therefore has the form $H=H_0+H_{DM}$, where the Dzyaloshinskii-Moriya interaction takes the form:
\beq
H_{DM}=-\sum_{{\bf r}} {\bf D} \cdot {\bf S}_{{\bf r}} \times ({\bf S}_{{\bf r}+\delta_1}+{\bf S}_{{\bf r}+\delta_2})
\label{MeasCritHam2}
\eeq
where the experimentally measured parameters are $J\sim 4.3$K, $J'/J \sim 0.34$ and
$|D|/J \sim 0.053$ (Ref.~\cite{coldea}c). The Dzyaloshinskii-Moriya interaction breaks the $SU(2)$ symmetry down to a $U(1)$ symmetry, called the ``easy-plane'' symmetry. 

The model presented cannot be solved analytically or
numerically except on very small lattices. Again due to the frustrated interactions, Monte Carlo
calculations are plagued by the sign problem. To gain an understanding of the physics in the particular
phases, it is instructive to consider a simpler model, which has the same
symmetries as the Hamiltonian in Eq. (\ref{MeasCritHam1}) and
(\ref{MeasCritHam2}). The fact that the Dzyaloshinskii-Moriya
interaction breaks the $SU(2)$ to $U(1)$ symmetry suggests to use an
anisotropic exchange interaction $J^z_{ij}<J_{ij}$, leading to the following
easy-plane spin model:
\beq
H_{{\rm xy}}=\sum_{\left< ij\right>} [J_{ij} (S^x_i S^x_j+S^y_i S^y_j)+J^z_{ij}S^z_i S^z_j],
\label{EasyPlaneHam}
\eeq 
where the notation $J_{ij}$ reminds us that there are two different
coupling strenghts $J$ and $J'$ in the real material Cs$_2$CuCl$_4$.

\subsection{Chain of transformations: from spins to bosons, from bosons to bosonic vortices and from
  bosonic to fermionic vortices}

Even this simpler model is still hard to solve. The theoretical challenge is
to find other variables, which correctly describe these spin systems. Here we
present one of the proposed approaches \cite{Alicea}, which allows one to
make progress, based on various transformations. In the limit $J^z_{ij}=0$, the
model in Eq.~(\ref{EasyPlaneHam}) is the quantum version of the $XY$ model. In
$2d$, we know that the classical $XY$ model can be equivalently described in terms of vortices. 
An analogous duality transformation in the quantum system will lead to the description of the system
in terms of {\it quantum vortices}. A second transformation, which fermionizes the
vortices, which are bosons, proves useful to make the system tractable. This
change of variables, called the Chern-Simons transformation~\cite{fradkin}, is the $2d$
generalization of the well-known Jordan-Wigner transform in $1d$.

First, we conveniently describe spin $1/2$ objects in terms of hard-core
bosons. A spin up on a given site corresponds to the site being occupied by
a boson, whereas a spin down corresponds to an unoccupied site. Hard-core
bosons means that there is at most one boson per site. Then the
$x-x$ and $y-y$ spin interactions are mapped onto the hopping of bosons,
which is described by creation $S^+_i=S^x_i+iS^y_i$ and annihilation
operators $S^-_i=S^x_i-iS^y_i$. On a bipartite lattice, the sign of
spin interactions (ferromagnetic or antiferromagnetic) is irrelevant, since
the sign of the hopping term can be easily removed with a simple unitary
transform. But for non-bipartite lattices, such as the triangular lattice,
antiferromagnetic interactions between the spins lead to a positive sign for
the hopping term.   Within a boson picture this does not make sense since the
bosons energy must be lowered by a near-neighbor hop. In order for the kinetic energy to have the ``correct" negative sign for the hopping
strength, one needs to do a canonical transformation to include a fictitious
flux of $\pi$ through the triangles of the lattice. When the boson hops
around a triangular plaquette, its phase picks up a factor of $\pi$,
reestablishing the sign due to antiferromagnetic interactions. This
$\pi$ flux produces a fictitious magnetic field, in which the boson hopping
takes place. The Hamiltonian Eq.(\ref{EasyPlaneHam}) in terms of bosons then takes the following
form 
\beq
H_{xy}=-\sum_{\left< ij\right>}[\frac{J_{ij}}{2} S^+_iS^-_j e^{iA^0_{ij}}+h.c.+J^z_{ij}S^z_iS^z_j],
\label{CritBosHam}
\eeq
where the last term is interpreted as a nearest neighbour
density-density interaction between the bosons and where $\int_{\rm
  triangle} {\bf A^0 \cdot {\bf dl} = \pi}$.

\begin{figure}
\centerline{\includegraphics[height=4cm,width=7.5cm]{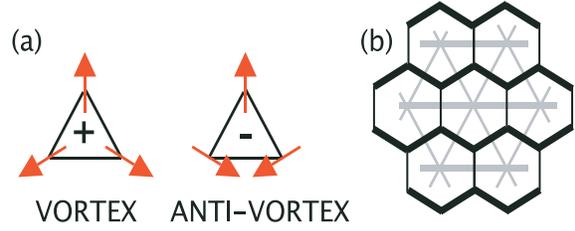}}
\caption{(a) A pictorial representation of vortices and antivortices. (b) The honeycomb lattice is dual to the triangular lattice.}
\label{fig:vortices}
\end{figure}

We now proceed with the next transformation. Bosons (in magnetic fields) are
often described in terms of {\it vortices}~\cite{fisherlee}. Let us now focus on
this new object and its properties. To define a vortex,
one can consider the phase of the boson creation operator 
$S^+_i\sim e^{i\phi}$ (or equivalently the
spin direction in the XY plane). A vortex is present when the integral of
the gradient of the phase $\phi$ around a loop gives $\int {\bf \nabla} \phi
\cdot {\bf dl}=2\pi$. Depending on the configurations of the spins in the triangle, they
may result in a vortex, associated with a vortex number $N=1$, or an
antivortex, associated with a vortex number $N=0$, as shown in Fig.~\ref{fig:vortices}a. 

Since the original spin problem is frustrated, the two
spin configurations in Fig.~\ref{fig:vortices}a are energetically equivalent.
Therefore there are as many triangular plaquettes with vortices as with
antivortices; the dual vortices are at {\it half-filling}.

In this description, the vortices live on the honeycomb lattice, which is dual to the triangular
lattice. This is easy to see, by connecting sites at the centers of
plaquettes of the triangular lattice - see Fig.~\ref{fig:vortices}b.  The duality transformation between bosons (the
original spins $1/2$) and vortices introduces a gauge field ${\bf a}$ sitting on
the links of the dual lattice. This gauge field will encode the long-range interaction
between the vortices.  This exact mapping relates, for example, the $z$
component of the spin to the field:
$S^z+1/2=\frac{1}{2\pi}|{\bf \nabla} \times {\bf a}|$. This indicates that a
vortex hopping around a dual plaquette will pick up a $\pi$ phase factor. This is easy to
understand as a vortex hopping around a plaquette of the honeycomb lattice
will encircle one spin $1/2$, and thus ``feel'' its average flux. Since on average
$\left< S^z\right> =0$ (the original spins are {\it not} in a magnetic field), the
vortices experience an average background magnetic field of flux $\int
d^2{\bf r} {\bf \nabla} \times {\bf a^0}=\pi$. ${\bf a^0}$ is the static part of
the gauge field, precisely designed to account for the fact that the
original spins can take only the values $S^z=\pm1/2$.

The gauge field actually mediates a long-range Coulomb, or
``electromagnetic'', interaction between
the vortices. In $2d$, the Coulomb potential is logarithmic. The Hamiltonian in the vortex
representation, where $b^{\dagger}$ and $b$ are the vortex creation and
annihilation operators, reads:
\begin{eqnarray}
H & = & -\sum_{\left< ij\right>} (t_{ij}b^{\dagger}_i b_j e^{i
  a_{ij}+a^0_{ij})}+h.c.) \nonumber \\
&~ & ~~~~~~+ \sum_{ij} (N_i-1/2)V_{ij}(N_j-1/2) \nonumber \\
&~ & ~~~~~~+ U\sum_r (\nabla \times {\bf a})_r^2.
\label{VorHam}
\end{eqnarray}

The first term is the kinetic
energy of the vortices, moving on a honeycomb lattice in a field with static component satisfying $(\nabla \times {\bf a^0})_r=\pi$ for all sites
r of the triangular lattice. The second term is the Coulomb interaction between
vortices with $V_{ij}\sim \ln(|i-j|)$. In this term we clearly see the
half-filled natures of the vortices, as $N_i=b^\dagger_ib_i$ is the vortex
number operator. The final term - similar to a
magnetic energy term - accounts for the dynamics of the gauge field.
We refer again to Ref.~\cite{Alicea,fisherlee} for a rigorous construction
of the dual vortex hamiltonian.

This type of boson/vortex transformation is especially useful
when the dual vortex variables are at integer filling. Here, however, due to the original spin frustration,
the vortices are located at half-filling, since the states with filling fraction of zero
or one are equally likely.  The dual vortex representation
thus appears as intractable as the original spin model.
However,
further progress can be made by performing yet another transformation,
``transmuting"  the bosonic vortices into {\it fermionic vortices}. The main
advantage of this change of variables, is that the interacting boson field
may be described using a non-interacting fermion field, since fermion
statistics have Pauli exclusion built in. This makes the model
mathematically easier. However, there is a price to pay. The fermion wavefunction changes sign upon
exchanging particles, whereas the boson function does not. The
transformation must therefore correct for this sign change. This Chern-Simons
transformation~\cite{fradkin} may be intuitively understood, as attaching a
fictitious $2 \pi$ flux tube to a fermion vortex in order to represent the
bosonic vortex. So if we consider the exchange interaction between two
fermionic vortices, due to the Aharonov-Bohm effect, encircling one particle
half way around the other results in a $e^{i \pi}$ flux from the interaction
of the fermion with the flux tube of the other particle, which cancels the
negative sign from fermion statistics.

Now let's look at a static picture of the fluxes attached to these
fermions, which also live on the dual lattice. We can use a convention where
each fermion shares on average its $2 \pi$ flux amongst the three honeycomb plaquettes
to which it belongs. Since on average, each honeycomb plaquette has three
vortices, as the vortices are at half-filling, it is therefore
pierced by a $2 \pi$ flux. This is equivalent to a zero flux and therefore irrelevant for the behaviour of the fermions. Of course, this is a crude picture, but it is valid in the mean field approximation where we ignore flux
fluctuations around their average. Therefore, the fluxes associated with the
Cherns-Simon transformation cancel to zero when we smear out these fluxes in
a mean field fashion. We note that any mean-field approximation is of
course crude, but here it is actually expected to be rather good,
since the interactions between the vortices tend to supress their density
fluctuations, and therefore the flux fluctuations in the Cherns-Simon
language. 

So now in the mean-field picture we are left with free fermions. However, we
have to remember that these fermions are {\it also} coupled to the original
gauge field ${\bf a}$, as were the vortices. In mean-field theory, we only consider the static contribution from
${\bf a^0}$, and we are thus describing a system of free fermions on the
honeycomb lattice with a background $\pi$ flux. This is a one-particle problem and it is easily
solved. The corresponding Fermi ``surface" consists of only four points, with
Dirac-like (gapless) dispersions around each one. 

The non-interacting limit we are currently describing, is a good starting point
to now re-introduce the gauge fluctuations we supressed in the mean-field
approximation. To do this, we now describe the behaviour of the low
energy particle-hole excitations around these four Fermi points in terms of
a Dirac theory for fermionized vortices. There are four flavours of Dirac
particles, which correspond to the four Fermi points. These fermionic
vortices interact logarithmically, which is accounted for by a coupling to
the gauge field ${\bf a}$. Additionally, they are coupled to the
Cherns-Simon gauge field ${\bf A_{\rm \bf CS}}$ due to the flux tube. It can be argued~\cite{Alicea}
that the field due to the flux tubes fluctuates much more than the field due
to the Coulomb interactions between vortices and that at long
length scales the Chern-Simons field can be ignored. This may
also be seen explicitly by defining an auxilary field and integrating out
the field due to the fictitious flux tubes. In fact, the logarithmic
interaction in $2d$ is so strong that it dominates over the statistics
encoded by the Cherns-Simon flux tubes. In this respect the description in
terms of fermions and bosons is analogous. We invite the interested reader
to consult Ref.~\cite{Alicea} for a more detailed discussion. 

The final formulation of the critical spin liquid in terms of $N=4$ flavours of
Dirac particles, which interact by a logarithmic potential in $2d$ is in
terms of a QED (Quantum Electrodynamics) Lagrangian in $2+1$ dimensions 
\beq
\it{L}_{QED3}=\bar{\psi_a} \gamma^{\mu}
(\partial_{\mu}-i\tilde{a_{\mu}})\psi_{a}+\frac{1}{2e^2}({\bf \nabla} \times
   {\bf \tilde{a}})^2,
\label{QED3}
\eeq 
where $\psi_a$ is the spinor associated to the Dirac fermion flavour $a\in
\{1,2,3,4\}$, $\gamma^{\mu}$ are Dirac gamma matrices in
direction $\mu \in \{x,y,z\}$ and ${\bf \tilde{a}=a+A_{\rm \bf CS}}$ is the
auxiliary field  used in the integration of the Cherns-Simon field ( the sums over
double indices are implicit). Strictly speaking, there should be an additional $\it{L_{4f}}$ four-fermion
term that originates from short-range interactions between the vortices that we have not taken into account
explicitly, but that can always be present in the microscopic model. It is
known that such terms are irrelevant in the limit of an infinite fermion
flavour number $N\rightarrow \infty$ and for all $N>N_c$,
where the exact value of $N_c$ is not known. 

Assuming $N_c<4$, we can safely
ignore this term. Now, we can finally import the results from QED, which
indicate that Eq.~\ref{QED3} describes a {\it critical theory} with {\it
  gapless excitations}. Thus as a result of these consecutive changes of variables, from spins to Dirac
particles, this approach describes the much desired critical spin liquid phase, without fine
tuning of variables and no free particles. This specific spin liquid is
dubbed an ``algebraic vortex liquid''~\cite{Alicea}.

In this section, we have tried to describe the different steps of the
construction that lead to the algebraic vortex liquid. We will now only
mention some of the properties of this type of spin liquid. The critical phase not only does
not break any symmetries, but a $SU(4)$ symmetry emerges. Above all, the
theory is able to make predictions, which may be compared with
experiment. The structure factor is predicted to have power law correlations
around five distinct momenta {\it with the same exponent for each wavevector}.
Two of these momenta correspond to the experimentally measured magnetic long range order wave vectors.
To make a further comparison with Cs$_2$CuCl$_4$, the Dzyaloshinskii-Moriya interaction
be incorporated, and in the Dirac theory corresponds to a ``mass" term which 
destabilizes the algebraic vortex
liquid driving it into an incommensurate spiral state at low enough temperature: this
is exactly what is observed in the experimental phase diagram of
Cs$_2$CuCl$_4$ (see Fig.~\ref{fig:expPD}a). 

The proposed algebraic vortex liquid state is not the only existing
scenario to describe the physics of Cs$_2$CuCl$_4$. There have been recent theoretical proposals 
such as a quasi-1d approach~\cite{tsvelik}, an algebraic spin liquid state
with fermionic spinons and a $U(1)$ compact gauge field~\cite{wen} or the
proximity of a quantum critical point between a $Z_2$ spin liquid and a
spiral phase~\cite{isakov}. 

We would finally like to emphasize that through the example
described in this section, we can see that, like in the topological spin liquid case,
the natural description of these phases is in terms of {\it gauge fields}.
The gauge fields that emerge for critical spin liquids, however, have
different symmetries (here $U(1)$) than the $Z_2$ symmetry for
the topological spin liquids of Sec.~\ref{sec:topo}.
We will further exploit gauge theories in the
next section.

\section{Deconfined quantum critical points}
\label{sec:dec}

We will now turn from the description of exotic phases
of matter discussed in the two previous sections, to an exotic {\it critical point}
separating two conventional states of matter. We will focus on the
quantum phase transition between a N\'eel state, which breaks spin rotational symmetry, and a Valence Bond Solid
state, which breaks translational symmetry, depicted in Fig.~\ref{fig:neel} and \ref{fig:vbs}b. We will show that, contrary to general
expectations of the Landau theory of phase transitions, 
a direct second order transition is possible. It is a continuous phase transition between two ordered states with different symmetries. We will find that, right at the critical
point, the spinons we encountered in the two previous sections are
essentially  {\it
 deconfined}, whereas they are altogether absent in both surrounding phases. The
description of this so-called ``deconfined quantum critical point'', will
also be in terms of a gauge theory, namely a non-compact $CP^1$ theory in this case.

\begin{figure*}
\centerline{\includegraphics[height=3.8cm,width=14cm]{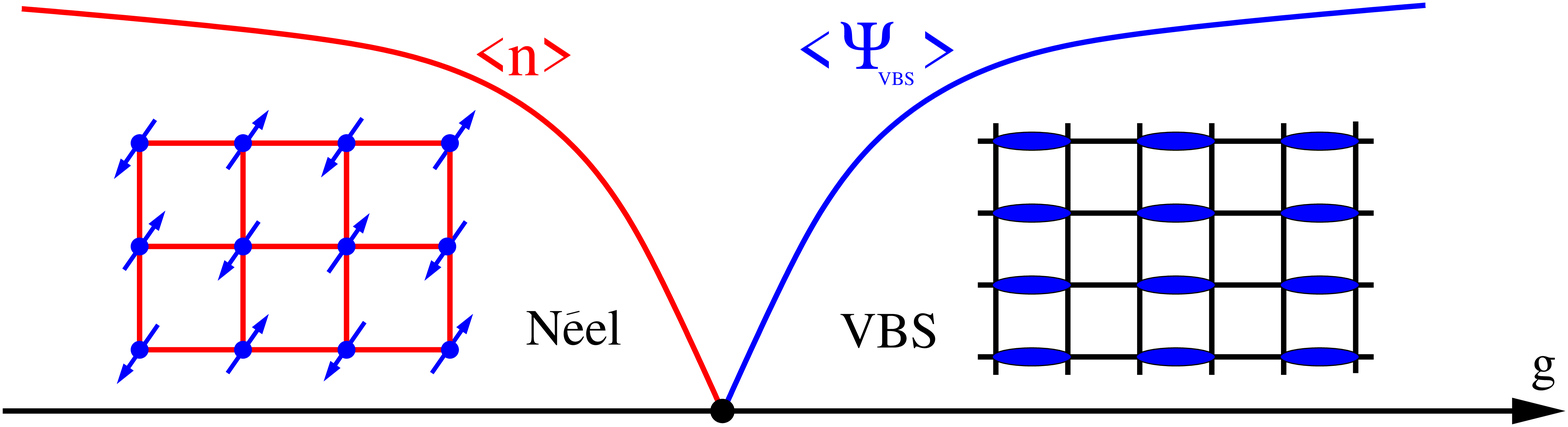}}
\caption{The $T=0$ transition (driven by an external parameter $g$) between
  a N\'eel antiferromagnetic order and a Valence Bond Solid for spins
  $S=1/2$ on the square lattice. The transition is argued to be
  second-order~\cite{senthil} with both N\'eel and VBS order
  parameters vanishing continuously at the critical point.}
\label{fig:neel2vbs}
\end{figure*}

\subsection{Formulation of the problem: the transition and standard Landau theory}

We want to describe $2d$ quantum $S=1/2$ spin systems with one spin per
unit cell on the square lattice and the quantum phase transition, {\it i.e.} a transition at
zero temperature driven by an external parameter, between a N\'eel state and
a VBS state (see Fig.~\ref{fig:neel2vbs}). We do not explicitely mention a specific form of
the microscopic interaction strength $g$ that drives the transition, which could be
longer-range couplings like in Hamiltonian Eq.~(\ref{HeisHam}), or maybe a ring-exchange term,
but rather focus on the symmetries of the associated phases. 

If instead of a VBS state a simple paramagnet with no symmetry breaking
existed on the right-hand side of Fig.~\ref{fig:neel2vbs},
it would be natural to describe the transition
within a standard Landau approach. That is, we
would identify a local order parameter, the staggered magnetization ${\bf
  n}_i=(-1)^{{\bf r}_i} {\bf S}_i$, coarse-grain it to the continuum
${\bf n}({\bf r})$ and expand the free energy or Lagrangian in gradients or
powers of ${\bf n}({\bf
  r})$ allowed by the symmetries such as $L({\bf n})=|{\bf \nabla n}|^2+r|{\bf n}|^2+u|{\bf
  n}|^4+\ldots$. Depending on the precise value of the parameters,
we would get either the N\'eel state ($r<0$) or the paramagnet ($r>0$), and
the transition at $r=0$. Here the main problem is that the plain paramagnet
{\it does not exist} in 2D at $T=0$, according to Hastings' theorem. It
is thus natural to consider something else than the paramagnet, more
specifically the VBS. In the case of a transition between two
different broken symmetry phases, the Landau theory would expand the
Lagrangian in terms of both the N\'eel order parameter ${\bf n}({\bf r})$
and the VBS order parameter $\psi_{\rm VBS}$ with possible cross terms.

For a phase transition between states with different broken symmetries, the
resulting Landau theory predicts one of the four possible
scenarios: (i) a direct first order phase transition, (ii) an intermediate
coexistence phase, between N\'eel and VBS orders in our case, (iii) an intermediate phase
with no order at all, (iv) a second order transition in the case of a
multi-critical point. Scenario (iii) can be excluded using
Hastings' theorem. Scenario (iv) is very unlikely, because it
is not generic and it implies a fine-tuning of some parameter.
Scenarios (i) and (ii) cannot be ruled out on
general grounds, but both exclude a direct continuous (second-order)
transition.

In contradiction to this standard Landau analysis, recent
work~\cite{senthil} shows that in the specific case of a N\'eel-VBS
transition on the square lattice, the transition {\it can generically be second
order} without any fine-tuning, due to subtle quantum interference effets that invalide the Landau
analysis. The crucial elements responsible for a continuous transition are topological
objects called {\it hedgehogs}. The resulting field theory is a non-compact gauge theory. In
what follows, we review in a simplified way the approach of
Ref.~\cite{senthil}.

\begin{figure*}
\centerline{\includegraphics[height=6.5cm,width=8.5cm]{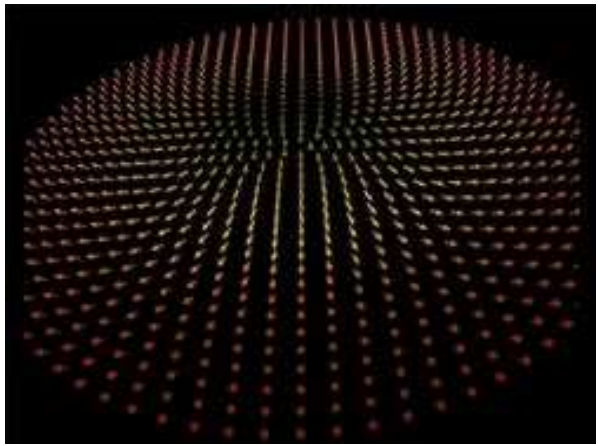}}
\caption{Real-space representation of a skyrmion in the N\'eel field ${\bf
    \tilde{n}}$ at a given timeslice. Spins in the center are pointing down whereas spins on the
    boundaries point up. The charge of this skyrmion is $Q=+1$.}
\label{fig:skyrmion}
\end{figure*}

\subsection{N\'eel order, skyrmions and hedgehogs}

\begin{figure*}
\centerline{\includegraphics[height=6.5cm,width=14cm]{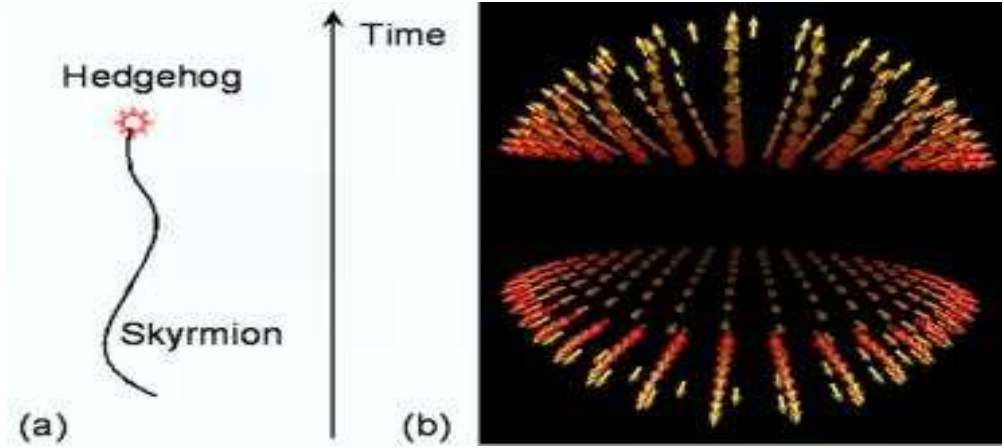}}
\caption{(a) The skyrmionic charge is suddenly changed at a {\it hedgehog}
  event in space time. (b) Real-space representation of a hedgehog event
  (the two spin configurations represent different time slices). A hedgehog
  corresponds to a singular configuration of ${\bf \tilde{n}}(x_\mu)$ at one
  space-time point where the skyrmionic charge changes.	All
  spins are pointing outwards of the hedgehog.}
\label{fig:hh}
\end{figure*}

To gain insight into such a N\'eel-VBS transition, let us first consider the
N\'eel state, and its topological defects. Since we are
considering a $2d$ quantum system at $T=0$, the N\'eel vector lives in
$(2+1)$ dimensions, where the extra-dimension can be interpreted as
imaginary time in
the standard path integral formalism. We shall denote the space-time coordinates as $x_\mu$ with $\mu \in
\{x,y,\tau\}$. Consider smooth configurations of the staggered magnetization ${\bf
  n}(x_\mu)=(-1)^{x+y} {\bf S}(x_\mu)$, {\it i.e.} ${\bf n}(x_\mu)$ is
slowly varying and is of constant amplitude. Let's define a unit
length N\'eel vector ${\bf  \tilde{n}}(x_\mu)={\bf  n}(x_\mu)/|{\bf
  n}(x_\mu)|$. For the N\'eel ground states, all the ${\bf  n}(x_\mu)$ are pointing in the same direction, but there exist configurations which admit
  a topological defect called a {\it skyrmion}. A skyrmion is represented in
  Fig.~\ref{fig:skyrmion}, in which the N\'eel vector at the center of the
  ``sample'' points down whereas it points up at the borders. Mathematically, the skyrmion number at a given time slice can be defined as an integer topological number
\beq
Q=\frac{1}{4\pi}\int dxdy ({\bf  \tilde{n}} \cdot \partial_x {\bf  \tilde{n}}
\times \partial_y {\bf  \tilde{n}})
\eeq
sometimes called an integer charge. If the spins were living in the
continuum, this number would be a conserved quantity in time. However, on a
lattice, this number can change from one time slice to the other. Such a
space-time event where the skyrmion number $Q$ changes by $\pm 1$ is called a
{\it hedgehog} (see Fig.~\ref{fig:hh}a). The choice of this name hopefully becomes apparent
upon inspecting
Fig.~\ref{fig:hh}b, where a hedgehog is represented. We can clearly see a
singular point from which all the ${\bf \tilde{n}}$ are pointing outwards,
like quills on a hedgehog. A hedgehog is the term used for a configuration of unit N\'eel vectors, which is singular at one space-time point, but smooth elsewhere. What is the role of these topological defects?  In a
N\'eel state, hedgehogs are very costly energetically and are therefore
absent. Deep within an ordinary paramagnet, the spins fluctuate essentially  independently
of one another.   In this case there is no energetic reason why hedgehog events should
be suppressed: we say that the hedgehogs proliferate. 

To see what happens to these hedgehogs on the
VBS side and at the critical point, it is useful to consider the non-linear
sigma model description of the N\'eel state~\cite{sigma} with the following action:
\beq
S_\sigma=\frac{1}{2g} \int dx_\mu |\partial_\mu {\bf  \tilde{n}}|^2 + S_{\rm Berry}
\eeq
where the first term accounts for the slow variations of ${\bf \tilde{n}}$,
$g$ is some ``stiffness'' constant, and the second term is a {\it Berry phase}
term which accounts for the quantum nature of the $S=1/2$ spins. Now let's
see the effects of hedgehogs on such a formulation by expanding the partition
function $Z=\int d{\bf  \tilde{n}} exp(-S_\sigma)$ in powers of the fugacity $\lambda$
of such events:
\begin{eqnarray}
Z & = & Z_0+\int_{x^1_\mu} \lambda (x^1_\mu) Z_1(x^1_\mu) \nonumber \\
 &  & ~~~ + \frac{1}{2}\int_{x^1_\mu,x^2_\mu} \lambda (x^1_\mu) \lambda (x^2_\mu)Z_2(x^1_\mu,x^1_\mu,)+\ldots
\label{eq:fug}
\end{eqnarray}

where $Z_0$ describes a model where no hedgehogs are allowed. It can be shown
(see Ref.~\cite{sigma}a) that due to the Berry phase terms on the square
lattice, the contribution $Z_1(x^1_\mu)$ from single hedgehog event is
oscillatory in space-time and thus rapidly vanishing upon integration
over $x^1_\mu$ - it is therefore strongly ``irrelevant".
The same is true for $Z_2$ and $Z_3$, and the first
non-vanishing contribution is from quadruple-hedgehog events ($\lambda_4\sim \lambda^4$) corresponding to skyrmion number changes of $\pm 4$. In powers
of the one-hedgehog fugacity, this is already a ``rare'' event, as it
requires four hedgehog events acting at the space-time point. It is
compellingly argued in Ref.~\cite{senthil} that $\lambda_4$ is
also {\it irrelevant} in a
renormalization group (RG) analysis right at the critical point. Therefore,
hedgehogs are completely absent, in a RG sense, at the phase transition.

Now what is the role of the hedgehogs in the VBS? Whereas they are
completely irrelevant in the N\'eel state, hedgehogs turn out to be crucial
for the onset of VBS order. It can be rigorously shown~\cite{read} that the skyrmion creation operator,
identified with the presence of a hedgehog, is equal to the VBS order parameter
$\psi_{\rm VBS}$ up to constant (see Ref.\cite{senthil}b for an intuitive
derivation). This essentially means that {\it when
hedgehogs proliferate, they do so in such a way that the spins pair 
into singlets and form a valence bond solid}. This is again due to Berry phase
effects. This dual property of hedgehogs - simultaneously destroying N\'eel and creating VBS orders
- cannot be captured within a standard Landau analysis. The emergence of VBS order is
moreover in agreement with Hastings' theorem which precludes a conventional
gapped paramagnet.

So, the non-linear sigma model, supplemented with an analysis
of the role of hedgehogs, contains in a way all the physics of both N\'eel and VBS
phases. However, it is not possible to build a Landau theory
out of it to describe the transition between both states, because hedgehogs
are absent right at the critical point as we argued above.

\subsection{Gauge theory description}

We now consider an alternate formulation 
which enables us to glean more insight into the physics of the
deconfined quantum critical point. Once again, gauge theory will provide the
natural framework.
We first rewrite the N\'eel vector in terms of spinor fields $z_\alpha$,
where $\alpha=\uparrow, \downarrow$,
\beq
\label{eq:spinor}
{\bf  \tilde{n}}=z^\dagger_\alpha {\bf \sigma}_{\alpha, \beta} z_\beta,
\eeq
where ${\bf \sigma}$ is the vector of Pauli matrices and the sum over double indices is implicit. Also,
the normalization $|{\bf  \tilde{n}}|^2=1$ requires $|z_\downarrow|^2+|z_\uparrow|^2=1$.
This is the so-called $CP^1$ representation of the $SU(2)$ algebra. The
spinors $z_\alpha$ can roughly be interpreted as the spinon fields.
Eq.~\ref{eq:spinor} simply means that we need two of these $S=1/2$ objects to get the $S=1$ spin-flip N\'eel field.

By definition, the spinor fields have a local $U(1)$ gauge redundancy -
if we multiply $z$ by some phase factor $\exp(i \gamma(x_\mu))$, ${\bf
  \tilde{n}}$ stays unchanged. The spinons are therefore coupled to a $U(1)$
gauge field $a_\mu=\Im (z^\dagger \partial_\mu z)$. The non-linear sigma
model action without Berry term can therefore equivalently be written as
\beq 
S_\sigma = \frac{1}{g} \int dx_\mu |(\partial_\mu-i a_\mu)z_\alpha|^2.
\eeq

Where are the hedgehogs in this formulation? Within the $CP^1$
representation, the skyrmion number $Q$ can be
shown to be simply related to the flux of the gauge field
\beq
Q = \frac{1}{2 \pi} \int dxdy (\partial_x a_y-\partial_y a_x).
\eeq
A skyrmion is interpreted as a flux charge, and the hedgehog which changes
this number is therefore a {\it monopole} in the gauge flux. A monopole
event is sometimes called an instanton. Precisely at
the critical point, the hedgehogs/monopoles have been argued to be
irrelevant and their absence gives rise to a supplementary conservation law:
the gauge flux is a conserved quantity. In absence of monopoles, a gauge theory is said
to be {\it non-compact}. Thus a non-compact $CP^1$ gauge theory is a likely
candidate to describe the physics of the critical point, since it
incorporates from the very beginning both the spin $1/2$ {\it and} the
absence of hedgehogs. A Landau theory in terms of both order parameters of
the N\'eel and VBS phases does not incorporate this crucial feature, and
thus also misses the important fact that the spinons, although absent in both phases, are the good ``emerging'' degrees of freedom 
to describe the system at the critical point.

\subsection{Non-compact $CP^1$ theory and phase diagram}

With these arguments in hand, the authors of Ref.~\cite{senthil} now
argue that the N\'eel-VBS quantum critical point can be described by a
non-compact $CP^1$ gauge theory with the following Lagrangian
\begin{eqnarray}
L_{{\rm NCCP1}} & = & |(\partial_\mu-i a_\mu)z_\alpha|^2 + r |z|^2 + u
  (|z|^2)^2 \nonumber \\
 & & + \kappa (\epsilon_{\mu \nu \lambda} \partial_\nu a_\lambda)^2,
\label{eq:NCCP1}
\end{eqnarray}

where $\epsilon_{\mu \nu \lambda}$ is the antisymmetric tensor, and we recognize
a magnetic-like energy in the last term. This is a Landau-like expansion, but in
term of the spinon fields, which can admit a second-order phase
transition, at least within a mean-field analysis.   Specifically, for $r<0$, the model describes 
the N\'eel state, since the spinon field acquires a non-zero expectation
value $\left< z_\alpha \right> \neq 0$ and thus $\left< {\bf
  \tilde{n}}\right> =\left< z^\dagger_\alpha \right> {\bf \sigma}_{\alpha,
  \beta} \left< z_\alpha \right> \neq 0$. For $r>0$, $\left< z_\alpha \right> = \left< {\bf
  \tilde{n}}\right> = 0$, and we have a paramagnetic state, also with no hedgehogs. This
hedgehog-free state is not an ordinary gapped quantum paramagnet.   Rather, it turns out to be a critical $U(1)$ spin liquid,
as described in Sec.~\ref{sec:crit}. For $r=0$, we have a second-order
transition between these two states: therefore the N\'eel-VBS critical point
is governed by the physics of a N\'eel-$U(1)$ spin liquid transition.

\begin{figure}
\centerline{\includegraphics[height=4.7cm,width=7cm]{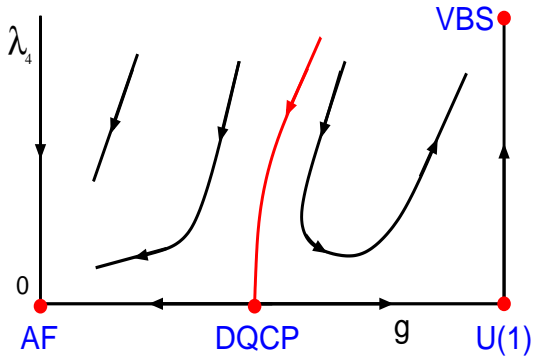}}
\caption{Renormalization-group flow phase diagram with parameters $\lambda_4$
  (the quadruple-hedgehog fugacity) and $g$ (the coupling strength that
  drives the transition). Four fixed points are visible (see text): the N\'eel
 (denoted AF), the VBS, the $U(1)$ spin liquid ($U(1)$ SL) and the deconfined 
quantum critical point (DQCP).}
\label{fig:RG}
\end{figure}

How can we gain intuition about this new phase transition?
Refs.~\cite{senthil,mw} argue that it can be described by a
simpler classical model: a 3d Heisenberg $O(3)$ model with no hedgehog
topological defects. Such a model is hard to define analytically; it
corresponds to the notation $Z_0$ in Eq.~(\ref{eq:fug}). However
on a computer, a Monte Carlo calculation can be performed by 
simulating a classical $O(3)$ magnet and rejecting ``by hand'' all
configurations which have a hedgehog. Such a calculation was
performed~\cite{mw} and a second order phase transition was found, between
the ferromagnetic state and the no-hedgehog paramagnet, with a different set of
critical exponents than if one allows free hedgehogs to exist. These calculations
showed that a description of the transition can be given in terms of the spinon
fields as in Eq.~(\ref{eq:NCCP1}), and not in terms of the N\'eel field. 
The observed transition is argued to be in the
same universality class as the N\'eel-VBS $2d$ quantum phase transition. 

The ``RG flow'' phase diagram in terms of the fugacity of the
quadruple-hedgehog event accounts for all the physical phases involved in the
above discussion (see Fig.~\ref{fig:RG}). The flow admits
four fixed points. The first point is the antiferromagnetic N\'eel phase, where the fugacity
renormalizes to zero, since the hedgehogs are too costly, and the VBS phase where
hedgehogs proliferate, since they create the VBS order. Then a novel
deconfined quantum critical point is present, where 4-hedgehogs events have been
argued to be irrelevant. Finally, the phase diagram admits a $U(1)$ spin
liquid point, which can be argued on general grounds~\cite{polyakov} to be
unstable with respect to the insertion of hedgehogs events. The deconfined quantum
critical point might be hard to observe in numerical calculations, because in a lattice model the ``bare'' $\lambda_4$
is in general non-zero. However, there are many experimental predictions that can be made
based on such a theory, in order to be able to detect deconfined criticality in
the laboratory. We refer the interested reader to references~\cite{senthil}
to find these predictions and to have a complete overview of deconfined
quantum criticality. Other pedagogical introductions to this topic are also available~\cite{senthil.simple}.

\section{Conclusion}
\label{sec:conc}

We have discussed how ``exotic'' behaviour can arise in condensed matter
systems which have strong interactions between electrons. Mott insulators
with one electron per unit cell are intrinsically strongly interacting, and a
good place to look for paradigm shifting phenomena. In particular, if no
symmetry is broken, the ground-state is guaranteed to be an exotic ``spin
liquid''. We have discussed two classes of spin liquids:
topological and critical. The former exhibit topological order and
gapped excitations with quantum number fractionalization. The latter display power-law
correlations in the ground state and gapless excitations. In both cases, gauge theory offers a simple way to characterize these spin
liquids and their excitations, and to encode the statistical interactions
between them. Duality transformations also let us gain understanding of
these novel quantum phases, which in some cases can be probed
experimentally, as in the spin $1/2$ compound Cs$_2$CuCl$_4$. 

We have also explored in the last section paths beyond Landau's theory of
phase transitions through the example of the deconfined quantum critical
point between a N\'eel and a Valence Bond Solid state in two dimensions. We
reviewed the recent theoretical work of Ref.~\cite{senthil} that proposes a
continuous transition between these phases, with emerging deconfined spinon
excitations right at the critical point. 

These recent developments hopefully convey the current excitement
in the field of strongly interacting quantum condensed physics.
The specific $Z_2$ and
$U(1)$ spin liquids that were discussed above surely represent only the tip of a large iceberg 
consisting of many other exotic
states of matters that can arise, especially in doped systems.  But much hard work
is needed to further understand and (hopefully!) detect such phases in the laboratory.

\section{Acknowledgments}
We first would like to thank the organizers of FPSPXI for the wonderful
summer school and the enjoyable atmosphere.
We would also like to thank Radu Coldea for kindly letting us reproduce the
experimental data of Fig.~\ref{fig:expPD}.
M.P.A.F. wishes to thank his many wonderful collaborators over the past ten years
(you know who you are!) as well as numerous other friends and colleagues for
many, many stimulating interactions (and for helping to keep him from straying
too far afield into the hinterlands of exotica) 
and acknowledges the 
National Science Foundation for generous support
through grants PHY-9907949 and 
DMR-0210790.
F.A. would also like to warmly thank Gr\'egoire Misguich for many discussions and
insightful comments on the manuscript.
A.M.W. wishes to thank {\L}ukasz Cywi\'nski for critical reading of the manuscript and the National Science Foundation through grants DMR0409937, PHY0216576 and PHY0225630 for support.

\end{document}